\documentclass[11pt]{article}
\usepackage{amsmath}
\usepackage{graphicx}
\usepackage{amsfonts}
\usepackage{amssymb}
\usepackage{epsf}
\usepackage{latexsym}

\def\80{\hspace{0.8in}}

\newcommand{\be}{\begin{enumerate}}
\newcommand{\ee}{\end{enumerate}}
\newcommand{\bi}{\begin{itemize}}
\newcommand{\ei}{\end{itemize}}
\newcommand{\bd}{\begin{description}}
\newcommand{\ed}{\end{description}}

\def\beq{\begin{equation}}
\def\eeq{\end{equation}}
\def\bea{\begin{eqnarray}}
\def\eea{\end{eqnarray}}
\def\foo{\footnote}
\def\pa{\partial}
\def\d{\textrm{d}}
\def\R{\underline{R}}
\def\r{\underline{r}}
\def\ttH{\mbox{\tt H}}
\def\ttL{\mbox{\tt L}}
\def\ttD{\mbox{\tt D}}
\def\ttP{\mbox{\tt P}}

\def\cr{\mbox{\scriptsize{\bf $\mbox{ } \times \mbox{ }$}}}

\def\sd{\mbox{\scriptsize d}}

\def\orb{\mbox{\scriptsize orb}}
\def\eff{\mbox{\scriptsize eff}}
\def\sC{\mbox{\scriptsize C}}
\def\sD{\mbox{\scriptsize D}}

\def\sR{\mbox{\scriptsize R}}

\def\sJacobi{\mbox{\scriptsize Jacobi}}

\def\eph(B){\mbox{\scriptsize emergent(LMB)}}

\def\uR{\mbox{\underline{R}}}
\def\sbfm{\mbox{{\bf \scriptsize\sffamily m}}}

\def\bfm{\mbox{{\bf \sffamily m}}}

\def\br{\mbox{{\bf r}}}
\def\bR{\mbox{{\bf R}}}
\def\bq{\mbox{{\bf q}}}
\def\bm{\mbox{{\bf m}}}
\def\bn{\mbox{{\bf n}}}
\def\sbm{\mbox{{\bf \scriptsize m}}}
\def\sbn{\mbox{{\bf \scriptsize n}}}

\def\sbA{\mbox{{\bf \scriptsize A}}}

\def\fE{\mbox{\sffamily E}}

\def\fI{\mbox{\sffamily I}}

\def\fL{\mbox{\sffamily L}}

\def\fT{\mbox{\sffamily T}}
\def\fU{\mbox{\sffamily U}}
\def\fV{\mbox{\sffamily V}}
\def\fW{\mbox{\sffamily W}}

\def\sfE{\mbox{\sffamily{\scriptsize E}}}

\def\sfT{\mbox{\sffamily{\scriptsize T}}}
\def\sfU{\mbox{\sffamily{\scriptsize U}}}

\def\b{\underline{b}}
\def\p{\underline{p}}
\def\q{\underline{q}}
\def\a{\underline{a}}

\renewcommand{\P}{\underline{P}}                 

\def\bn{\mbox{\bf n}}
\def\bp{\mbox{\bf p}}
\def\bA{\mbox{\bf A}}

\def\bP{\mbox{\bf P}}

\begin{document}
\begin{titlepage}
\vspace{.7in}
\begin{center}

\LARGE{\bf CLASSICAL DYNAMICS ON TRIANGLELAND}

\vspace{.4in}

\large{\bf Edward Anderson$^*$}

\vspace{.2in}

\large{\em Peterhouse, Cambridge CB2 1RD}\normalsize

\vspace{.1in}

\large{and}

\vspace{.1in}

\large{\em DAMTP, Centre for Mathematical Sciences, Wilberforce Road, Cambridge CB3 OWA.}

\end{center}

\begin{abstract}

In Euclidean relational particle mechanics (ERPM) only relative times, relative angles and relative 
separations are meaningful, while in similarity relational particle mechanics (SRPM) 
only relative times, relative angles and ratios of relative separations are.  
These theories are clearly of interest in the absolute or relative motion debate.  
In this paper, ERPM and SRPM are provided in fully reduced form for 3 particles in 2D, i.e. 
the classical dynamics on triangleland in 2D with and without scale.    
Exact solutions to each of these are then found, and simple Newton--Coulomb-like and harmonic 
oscillator-like SRPM models are studied numerically.  
The mathematics one arrives at thus overlaps in many ways with that which arises in the absolutist 
approach.  
The ERPM gives standard mathematics, while the SRPM has standard small-relative-scale behaviour and an 
unexpected but in itself standard universal large-relative-scale behaviour.  
One way in which SRPM is unusual is that it is a model in which a symmetry principle underlies an 
unexpected departure from standard physical behaviour at sufficiently large relative scales (interpolation 
between the abovementioned two behaviours).     
ERPM and SRPM are also theoretically interesting at the quantum level, both on their own merit 
and as toy models for the development of various approaches to the problems of time and of observables 
in quantum general relativity.  

\end{abstract}

\noindent PACS: 04.60Kz, 04.20.Fy.  


\mbox{ }

\vspace{4in}

\noindent$^*$ ea212@cam.ac.uk

\end{titlepage}

\section{Introduction}

The absolute or relative motion debate has been running for over three centuries \cite{Principia, LCC, 
Berkeley, Mach, DOD, buckets, JBook}.  
At the level of particle mechanics, those arguing for relative motion were until quite recently 
hampered by a lack of explicit examples of mechanics in which only relative times, relative separations 
and relative angles are meaningful.
These I call `relational' theories to avoid confusion with the standard conception of the theory of 
relativity.

%
%

Barbour--Bertotti theory \cite{BB82} is a such.  
It implements temporal relationalism directly and without extraneous coordinates 
by using a reparametrization-invariant Jacobi-type \cite{Lanczos} action.
It also implements spatial relationalism by some indirect means that involves auxiliary variables that 
represent arbitrary frame corrections with respect to Eucl(D) the D-dimensional Euclidean group of 
translations, Tr(D), and rotations, Rot(D), which is why I term this theory Euclidean relational 
particle mechanics (ERPM).  
This theory's implementation of spatial relationalism generalizes to more general theories of 
configurations for which a group $G$ of transformations to be rendered physically irrelevant 
\cite{BB82, Lan}.  
The corrections are in each case with respect to auxiliary variables that represent the infinitesimal 
generators of $G$, and in each case the configurational relationalism with respect to $G$ gets implemented 
through variation with respect to these auxiliaries producing constraints linear in the momenta which 
are constraints associated with $G$ and which use up both the introduced degrees of freedom (d.o.f's) 
and an equal number of d.o.f's of the absolute configurations.  
In the original paper \cite{BB82}, these auxiliary variables take the form of (`best matching')  
coordinates which correct the particle velocities, $\dot{\underline{q}}_I \longrightarrow 
\dot{\underline{q}}_I - \underline{k} - \underline{\omega} \cr \underline{q}_I$; here $\underline{k}$ 
is the translational auxiliary and $\underline{\omega}$ is the rotational auxiliary.   
However, this would spoil the aforementioned reparametrization invariance, so one would prefer 
these to be explicitly velocities, $\dot{\underline{q}}_I - \dot{\underline{a}} - 
\dot{\underline{b}} \cr \underline{q}_I$.   
This can be arranged prior to the variation by considering `corrected coordinates' 
\cite{ABFO, Lan, 06I, Phan}
${\underline{q}}_I \longrightarrow {\underline{q}}_I - \underline{a} - \underline{b} \cr \underline{q}_I$ 
and finding both that these only contribute to the action as the above auxiliary velocity corrections 
to the particle velocities and that the subsequent variation is still consistent and equivalent to 
the original one (see footnote 11).    

As well as playing a role in the history and philosophy of science literature (see e.g. 
\cite{Comments, buckets, B9599, EOT}), Barbour--Bertotti theory has also been geometrized, via the 
auxiliary variables turning out to be eliminable, as indicated in \cite{LB, Gergely}, and in \cite{06I} 
using the convenient Jacobi coordinates \cite{Marchal}.  
It was also noted in \cite{06II} that this elimination has a simpler nature in 2D than in 3D.    
But a nonredundant description for 2D in terms of relative separation and relative angle variables was 
not provided there.  
The first result of the present paper (Sec 2) is the 3-particle case of this, called `triangle land' 
in Barbour's popular science book \cite{EOT}, is recast as a system for two relative interparticle (cluster) 
separations and one interparticle (cluster) separation that takes an analogous mathematical form to the 
mechanics of 2 particles in 2D with a `shared' angular part.  
See \cite{FORD, RelatedWork} for further geometrical results.  
Barbour--Bertotti theory has also provided guidance in investigations of alternative conceptual 
foundations for GR \cite{BB82, B94I, RWR, Phan}.  
Moreover, most of the comments at the end of this Section about relational particle models serving 
as toy models for various approaches to quantization and to the problem of time in quantum GR to date 
concern, in particular, Barbour--Bertotti theory.

Barbour has additionally formulated a similarity relational particle theory (SRPM) \cite{B03}.   
This can be constructed by arbitrary frame correcting with respect to the D-dimensional 
group Sim(D) which now comprises the dilations as well as the translations and rotations.    
I have shown how this formulation's auxiliary variables are also eliminable \cite{06II}, 
thereby geometrizing it.    
The second result of this paper (Sec 4) is to extend this also to a nonredundant description 
for 3 particles in 2D, in terms of one ratio of separations and one relative angle, i.e. to a `triangle 
land' in which shape but not size is meaningful.   
This takes an analogous form to the mechanics of 1 particle in 2D, with these two 
coordinates forming a new (conformally) flat polar coordinate pair.
\cite{FORD} contains further geometrical results about SRPM, based on an earlier study in a 
somewhat different context \cite{Shape}.     
SRPM has also provided some guidance in investigations of alternative conceptual foundations for GR 
\cite{ABFKO} (this investigation also yielded a logical alternative to GR \cite{ABFO}).    
SRPM has only recently begun to be quantized and considered as a toy model for the Problem of Time 
\cite{06II, 07II}.

For both ERPM and SRPM, relative angle independence in the potential is a simplifying feature in analogy 
with how central potentials simplify ordinary mechanics.  
One should note that the scale-invariant problem differs from the ordinary central force problem in that 
it has restricted and unusual potentials inherited from the scale invariance.     
In Sec 3 I consider simple relational Newton--Coulomb and harmonic oscillator (HO) potentials 
(the 3 body problem within a plane suffices to make contact with 
many physically relevant situations at the classical level).   
In Sec 5 I consider the counterparts of these in SRPM, which are not Newton--Coulomb and HO  
per se, but do mimic these well for a wide range of situations in which one's subsystem 
separation is much smaller than its separation from an external massive object, although, interestingly,  
when it is no longer much smaller, deviations from the standard mechanics occur.
Investigation of the relative angle dependent cases, formulated in Sec 6, will be in a further paper 
\cite{AngleDep}.      
These workings provide further examples of dynamical orbits in triangle land, and, undoing some 
coordinate transformations, of paths of individual particles in the na\"{\i}ve position space 
(c.f. Figs 9-10 and 12-14 of \cite{EOT}).

Another forthcoming paper \cite{07II} studies this paper's HO-like\foo{The 2D Newton--Coulomb model has 
less motivation at the quantum level, as in the 3-d situation there, even though the classical motion is 
planar, there would usually be wavefunction spread in the third dimension.} SRPM model at the quantum 
level \cite{07II}.  
Via this, the present paper is further motivated by 1) the interesting question of whether relational 
physics is suggestive of any differences in quantum mechanical behaviour.  
2) At the level of gravitational theory, one issue at stake is whether GR succeeds in encapsulating the 
heart of relational/Machian thinking \cite{BB82, B94I, B9599}. 
Einstein made only an indirect approach at this \cite{Einstein}, while spacetime might be viewed as 
possessing some residual properties of absolutism.  
However, putting dynamics to the fore and emphasizing the configurations (3-geometries) as primary 
rather than spacetime, (globally hyperbolic, compact without boundary) GR can be demonstrated to have 
direct counterparts of the abovementioned spatial and temporal relationalism \cite{RWR, +RWR, Phan}. 
Then, given this close parallel and the lack of progress in quantum GR itself, 
relational particle models serve as one kind of toy model for addressing conceptual issues such as the 
problem of time (see e.g. the reviews \cite{Kuchar92, Kieferbook}) and the problem of observables, 
as well as various technical issues.
In particular, they have been used in Dirac quantization \cite{Smolin, 06I}, in group-theoretic 
quantization \cite{Rovelli}, in discussion of the semiclassical approach to the problem of time 
\cite{BS, 06I, 06II, SemiclI, SemiclII, SemiclIII}, of the timeless records approach 
\cite{B94I, B94II, EOT, 06II, Records} and of internal time approaches \cite{Paris, 06II, SemiclI}.  
There is related work in \cite{DeWitt} (conceptualizing about GR's configuration space) and in \cite{BY}.     
As explicit temporally relational models with nontrivial linear constraints, this paper's models 
are a further specific place to extend these studies (see the Conclusion for more details).

\section{Full reduction for 3-particle 2-d ERPM}

Begin in dimension D for $N$ particles, i.e. with $N$D absolute coordinates or $n$D relative position 
coordinates for $n = N - 1$.  
Consider an action that is closely related\foo{See footnote 11 for explanation of my precise choice of 
action.} 
to the original Barbour--Bertotti 1982 action \cite{BB82},\foo{{\bf Notation}:
\noindent I denote (absolute) particle position coordinates by $\bq = \{\underline{q}_I, I = 1, 2, 3\}$.  
These are defined with respect to a fixed origin and fixed coordinate axes.  
I denote relative particle position coordinates by $\br = \{\underline{r}_{IJ}, I > J\}$.  
No fixed origin enters their definition, but they are still defined with respect to fixed coordinate axes.    
I denote relative Jacobi coordinates by $\bR = \{\underline{R}_i, i = 1, 2\}$.  
These are a recoordinatization of a basis set of $\underline{r}_{IJ}$ that diagonalize the kinetic term.  
In the case of 3 particles, there is a unique prescription for these up to particle label permutations: 
$\R_1 = \q_2 - \q_3$ and $\R_2 = \q_1 - \frac{m_2\q_2 + m_3\q_3}{m_2 + m_3}$. 
I find it convenient to re-express these in bipolar form, which I denote by 
$\{\mbox{\boldmath$\rho$}, \mbox{\boldmath$\theta$}\} = \{\rho_i, \theta_i, i = 1, 2\}$.
One can now pass to fully relational coordinates $\mbox{\boldmath${\cal R}$} = \{\rho_1, \rho_2, \Phi\}$ 
for $\Phi$ = arccos$\left(\frac{\R_1\cdot \R_2}{||\R_1|| ||\R_2||}\right)$ the relational `Swiss army 
knife angle'.      


The particle masses are $m_I$. 
The Jacobi interparticle (cluster) reduced masses $\mu_i$ that feature in the diagonal kinetic term are 
then given by $\mu_1 = \frac{m_2m_3}{m_2 + m_3}$ and $\mu_2 = \frac{m_1\{m_2 + m_3\}}{m_1 + m_2 + m_3}$.  
Denote the inner product with respect to the array $\bA$ by 
$\mbox{}_{\sbA}(\mbox{ } \cdot \mbox{ })$, with corresponding norm $\mbox{}_{\sbA}|| \mbox{ } ||$. 
In particular, $\bm$ is the mass matrix with components diag($m_I$) in the $\underline{q}_I$ coordinate 
system and $\mbox{\boldmath$\mu$}$ is the relative Jacobi mass matrix with components 
diag($\mu_i$) in the $\underline{R}_i$ coordinate system.  
I also use  
$M = \sum_{I = 1}^Nm_I$ for the total mass, 
$I = \mbox{}_{\mbox{\scriptsize\boldmath$\mu$}} ||\bR||^2$ for the barycentric moment of inertia, 
$T = \mbox{}_{\mbox{\scriptsize\boldmath$\mu$}} ||\dot{{\bR}}   ||^2$ for the na\"{i}ve dikinetic energy,
$P = \mbox{}_{\mbox{\scriptsize\boldmath$\mu$}} ||\dot{{\mbox{\boldmath$\rho$}}}||^2$ for  
the na\"{i}ve radial dikinetic energy, 
$E = \mbox{}_{\mbox{\scriptsize\boldmath$\mu$}} ( \bR \cdot \dot{\bR})$ for the `Euler quantity' 
(or `dilational momentum'), and $I_i, T_i, P_i, E_i$ for the partial counterparts of these quantities.}    
\beq
\fI_{\sJacobi}[\q_I, \dot{\q}_I, \dot{\a}, \dot{\b}] = 
\int\d\lambda \fL_{\sJacobi}(\q_I, \dot{\q}_I, \dot{\a}, \dot{\b}) = 
2\int\d\lambda\sqrt{\fT\{\fE + \fU\}}
\label{A1} 
\mbox{ } .   
\eeq 
Here, temporal relationalism is implemented through the action being a Jacobi-type \cite{Lanczos} 
reparametrization invariant action.  
\beq
2\fT(\dot{\q}_I, \dot{\a}, \dot{\b}) = \mbox{}_{\sbm}\|\vec{{\cal E}}\bq\|^2
\eeq
is the dikinetic energy, where 
\beq 
\vec{{\cal E}}\q_I \equiv \dot{\q}_{I} - \dot{\a} - \dot{\b} \cr \q_{I}
\eeq 
are the particle velocities with Euclidean frame corrections, 
$\underline{a}$ being the translational auxiliary and $\underline{b}$ the rotational auxiliary.  
Such use of corrected frame implements the corresponding spatial relationalism via the 
variational procedure given at the start of Appendix A.  
While the above is written in a 3D-like form with a rotation vector $\underline{b}$, it also 
encapsulates the 1D case for $\b = 0$ and the 2D case for $\b = (0, 0, b)$, the third component 
being the one perpendicular to the 2D plane.
Finally, $\fU$ is the negative of the potential energy $\fV$, taken to be of the time-independent and 
relational form $\fV = \fV(||\br|| \mbox{ alone})$, and  $\fE$ is the total energy of the model 
universe.

\mbox{ }

\noindent[{\bf Figure 1 Caption.} Coordinate systems for 3 particles in 2D.    

\noindent i) Absolute particle position coordinates 
$\underline{q}_1$, $\underline{q}_2$, $\underline{q}_3$ with respect to fixed axes and a fixed origin O.

\noindent ii) Relative particle position coordinates, any 2 of which form a basis.     
 
\noindent iii) Relative Jacobi coordinates $\underline{R}_1$, $\underline{R}_2$.


\noindent iv) Bipolar relative Jacobi coordinates $\rho_1$, $\theta_1$, $\rho_2$, $\theta_2$.  
These are still with respect to fixed axes.  

\noindent v) Fully relational coordinates $\mbox{\boldmath${\cal R}$} = \{\rho_1, \rho_2, \Phi\}$.] 

\mbox{ }

I use the Lagrangian form of the constraints which follow from this action (Appendix A1) to 
eliminate the auxiliaries $\dot{\underline{a}}$ and $\dot{\underline{b}}$ (Appendix A.2).
This gives, in relative Jacobi coordinates, 
\beq
\fI_{\sJacobi}[\R_i, \dot{\R}_i] = 2\int\d\lambda \fL_{\sJacobi}(\R_i, \dot{\R}_i) = 
2\int \d\lambda\sqrt{\fT\{\fE - \fV\}}
\label{ARot}
\eeq
for
\beq
\fT(\R_i, \dot{\R}_i) = T/2 + \fT_{\sR} \mbox{ } , 
\label{fT}
\eeq
and 
\beq
2\fT_{\sR} = - \mbox{}_{\mbox{\scriptsize\boldmath${\cal I}$}}\|\mbox{\boldmath${\cal L}$}\|^2
\eeq
which, in 2D, is also 
\beq
- \sum_{i = 1}^n\sum_{j = 1}^n \frac{\mu_i\mu_j}{I}
\{(\R_i\cdot\R_j)(\dot{\R_i}\cdot\dot{\R_j}) - (\R_i\cdot\dot{\R_j})(\dot{\R_i}\cdot\R_j)  \} 
\mbox{ } 
\label{2Deasier}
\eeq
by virtue of the inertia tensor ${\cal I}$ reducing to just a number in 2D and by using the Kronecker $\delta$ theorem on 
the $\epsilon$ tensors contained in the barycentric angular momenta $\mbox{\boldmath${\cal L}$}$.
Note that while this {\sl expression} is manifestly independent of absolute angles because it is built 
solely out of dot products between relative vectors, but it is still a {\sl redundant description}: 
it uses $2n$ coordinates but there are only $2n - 1$ d.o.f.'s: the $\R_i$ variables themselves contain a 
vestige of reference to absolute orientation.

In this paper, I concentrate on the thus simpler 2D case. 
Applying the coordinate transformation and elimination of Appendix A.3, I obtain the relational non-redundant 
(and so, fully reduced) action 
\beq
\fI_{\sJacobi}[{\mbox{\boldmath${\cal R}$}}, \dot{{\mbox{\boldmath${\cal R}$}}}] = 
2\int\d\lambda\sqrt{\frac{1}{2}\mbox{ }_{\mbox{\scriptsize{\boldmath${\cal M}$}}}\|
\dot{\mbox{{\boldmath${\cal R}$}}}\|^2 \{\fE + \fU(\dot{{\mbox{\boldmath${\cal R}$}}})\}} 
\mbox{ } ,  
\label{ERPMac1}
\eeq
where ${\mbox{\boldmath${\cal M}$}}({\mbox{\boldmath${\cal R}$}})$ is the `mass matrix', with components  
$\mbox{diag}(\mu_1, \mu_2, \mu_3(\rho_1, \rho_2))$ in the ${\mbox{\boldmath${\cal R}$}} = 
\{\rho_1, \rho_2, \Phi\}$ coordinate system, that plays here the r\^{o}le of Jacobi--Synge dynamical 
metric \cite{Lanczos}.  The last entry is the configuration-dependent `mass',  
\beq
\mu_3 = \frac{I_1I_2}{I} = \frac{\mu_1\mu_2\rho_1^2\rho_2^2}{\mu_1\rho^2 + \mu_2\rho^2} \mbox{ } .  
\eeq
%
%
This configuration space geometry is is curved, e.g. its Ricci scalar is $6/I$.   
This configuration space geometry is useful at the classical level in this paper and at the 
quantum level in \cite{07II}.  
Or, explicitly in terms of the specific relational coordinates, this action fully reduced action is 
\beq
\fI_{\sJacobi}[\rho_1, \rho_2, \Phi, \dot{\rho}_1, \dot{\rho}_2, \dot{\Phi}] = 
2\int \d\lambda\sqrt{        \frac{1}{2}  
\left\{
\mu_1\dot{\rho}_1^2 + \mu_2\dot{\rho}_2^2 + 
\frac{    \mu_1\mu_2\rho_1^2\rho_2^2\dot{\Phi}^2    }{    \mu_1\rho_1^2 + \mu_2\rho_2^2    }
\right\} 
\{    \fE + \fU(\rho_1, \rho_2, \Phi)    \}          } 
\mbox{ } .  
\eeq

Appendix A.4 provides the variational equations that follow from this fully reduced action.  
In particular, the equations simplify (Appendix A.5) for $\Phi$-free potentials, which is the case for a 
number of relevant power-law potentials and their concatenations.  
What happens in this case is that $\Phi$ is a cyclic coordinate, giving a first integral in 
close analogy with the situation for a single particle in a central potential, in which case 
the conserved quantity is angular momentum.     
In the present context, what one has is two centrifugal terms with {\sl shared} angular momentum, 
corresponding to one subsystem having angular momentum $J$ and the other having angular momentum $-J$, 
so that overall there is indeed the zero angular momentum that the relational particle model requires.  
One can thus interpret the present context's conserved quantity $J$ as a 
`relative angular momentum quantity'.
This paper's ERPM solution examples (Sec 3) are then drawn from this simplified case.

The Hamilton--Jacobi formulation that follows from this action, on which the next Section's finding of 
solutions is based, is as follows.  
In the case of complete separability (Appendix A.6), Hamilton's characteristic function is  
\beq
\fW(\mbox{\boldmath${\cal R}$}) = \sum_{i = 1}^2 \int \d\rho_i 
\sqrt{2\mu_i\{\alpha_i + \fU(\rho_i)\} - \alpha_{\Phi}^2/\rho_i^2} + \alpha_{\Phi}\Phi \mbox{ } .  
\label{HJ1}
\eeq
Also,  
\beq
\Phi - \sum_{i = 1}^2 \int \frac{\mu_i\d\rho_i} 
{\rho_i\sqrt{2\mu_i\{\alpha_i + \fU(\rho_i)\} - \alpha_{\Phi}^2/\rho_i^2}} + 
\alpha_{\Phi}\Phi =
\pa_{\Phi} \fW = \beta_{\Phi} \mbox{ } , \mbox{ constant }
\label{6}
\eeq
is the equation of the relative orbits of the $\R_1$ and $\R_2$ subsystems.  
Denote the two integrals in the above equation by $L_1$ and $L_2$.  
Finally, 
\beq
\int \frac{\mu_i\d\rho_i} 
{\sqrt{2\mu_i\{\alpha_i + \fU(\rho_i)\} - \alpha_{\Phi}^2/\rho_i^2}} = \pa\alpha_i\fW = 
\tau + \beta_{\rho_i} 
\label{HJ3}
\eeq
are the orbit traversals in terms of a parameter $\tau$.  
$\tau$ is trivially eliminable from the i = 1 and i = 2 versions of the above equation: 
denoting the above integral by $K_i$,    
\beq
K_1 - K_2 =  
\int \frac{\mu_1\d\rho_1}{\sqrt{2\mu_1\{\alpha_1 + \fU(\rho_1)\} - \alpha_{\Phi}^2/\rho_1^2}} - 
\int \frac{\mu_2\d\rho_2}{\sqrt{2\mu_2\{\alpha_2 + \fU(\rho_2)\} - \alpha_{\Phi}^2/\rho_2^2}} = 
h \mbox{ } , \mbox{ constant}.
\label{8}
\eeq
Then the shape of the path in configuration space is given by (\ref{6}, \ref{8}) with the identifications 
$\alpha_{\Phi} = J$, $\alpha_i = \fE_i$.

\section{Some exact solutions of ERPM}

\subsection{Some physically interesting potentials}

Many potentials in physics are proportional to some power of the separation between two particles, 
$k_{IJ}||\underline{q}_I - \underline{q}_J||^{\alpha}$, or are linear combinations of these.  
In the present case of 3 particles, I use $k_1$ as shorthand for $k_{23}$ etc.  
I also replace some of the $k$'s with special labelling letters in the below examples.

In my relational coordinates, $||\q_2 - \q_3||^{\alpha}$ is just $\rho_1^{\alpha}$, while  
$||\q_1 - \q_3||^{\alpha}$ and $||\q_1 - \q_2||^{\alpha}$ are more complicated $\Phi$-dependent functions.   
As examples of potentials in the above class, 

\noindent 1) the 3 Newton--Coulomb potential, $\fV = - 
\frac{n_{23}}{||\underline{q}_2 - \underline{q}_3||}$ + cycles, becomes, in my relational coordinates,   
\beq
\fV = \frac{N_{3}}{\sqrt{\rho_2^2 + 2U\rho_1\rho_2\mbox{cos}\Phi + U^2\rho_1^2}}  
+ 
\frac{N_{2}}{\sqrt{\rho_2^2 + 2V\rho_1\rho_2\mbox{cos}\Phi + V^2\rho_1^2}} +  
\frac{N_{1}}{\rho_1} \mbox{ } ,    
\eeq
where $N_1 = n_{23}$ and cycles and I have introduced the dimensionless constants 
\beq
U \equiv  - \frac{2m_3}{m_2 + m_3} \mbox{ } , \mbox{ }
V \equiv  \frac{2m_2}{m_2 + m_3} \mbox{ } . \mbox{ }
\eeq
This is much simpler if only the 23 interaction is non-negligible, $\fV = N_{1}/{\rho_1}$.  

\noindent 2) The 3 HO potential, $\fV = h_{23}||\q_2 - q_3||^2 +$ cycles, 
becomes, in my relational coordinates,  
\beq
\fV = H_1\rho_1^2 + G\rho_1\rho_2\mbox{cos}\Phi + H_2\rho_2^2 
\eeq
where the effective Hooke's law constant coefficients are given by 
\beq
H_1 = h_{23} + \frac{h_{13}m_2^2 + h_{12}m_3^2}{\{h_{12} + h_{13}\}^2} 
\mbox{ } , \mbox{ }
H_2 = h_{12} + h_{13} 
\mbox{ } , \mbox{ }
\eeq
and the cross-term's constant coefficient is given by 
\beq
G = \frac{2\{h_{13}m_2 - h_{12}m_3\}}{m_2 + m_3} 
\mbox{ } .
\eeq
As well as the obvious simple subcase $\fV = H_1\rho_1^2$, the above has a wider simple subcase: if the 
original Hooke coefficients are chosen such that $m_2h_{13} = m_3h_{12}$, then the $\Phi$-dependence drops out 
and the potential becomes separable,  
\beq
\fV = H_1\rho_1^2 + H_2\rho_2^2  \mbox{ } .  
\eeq
I refer to this case as the `special multiple HO'.  
Its physical meaning is that the resultant force of the second and third `springs' points along 
the line joining the centre of mass of particles 2 and 3 to the position of particle 1.  
One virtue of this model is that, unlike the previous simple examples above, its potential is bounded, 
which is a useful feature in the quantum sequel of this paper.

\subsection{Simple exact solutions for $\Phi$-free potentials}

Each of the free-free, attractive Newton--Coulomb-free, HO-free and the  
aforementioned special multiple HO setting problems separate into single-variable problems.  
These amount to solving for the corresponding $L_i$, $K_i$, which themslves are standard computations:
Thus, overall, one can assemble solutions to each of the above ERPM problems from standard results.  

\mbox{ }

\noindent{\bf Zero relative angular momentum}

\mbox{ } 

\noindent In this special case the motion is linear (and indeed equivalent to the 1D problem at the 
classical level).
Then only the $K_i$ are needed.  
Denoting these by $K_i(\fU(\rho_i)) = K_i(k_i\rho_i^{\alpha}) \equiv K_i^{(\alpha)}$ 
and dropping the $i's$ on the $K_i^{(\alpha)}$, $\mu_i$, $\fE_i$, $k_i$, $N_i$, $H_i$ and $\rho_i$, 
the requisite $K^{(\alpha)}$ are 
\beq
K^{(0)} = \frac{\mu}{2\{\fE + k\}}\rho \mbox{ } ,
\eeq
\beq
K^{(-1)} = \sqrt{\frac{\mu}{2}}
\left\{
\frac{\sqrt{\rho\{\fE\rho + N\}}}{2\fE} - \frac{N}{\fE^{3/2}}\mbox{ln}
\left(\frac{N}{2\sqrt{\fE}} + \sqrt{\fE}\rho + \sqrt{\rho\{\fE\rho + N\}}
\right)
\right\} \mbox{ } ,  
\eeq
\beq
K^{(2)} = \sqrt{\frac{\mu}{2H}}\mbox{ln}\left(\sqrt{H}\rho + \sqrt{\fE + H\rho^2}\right) 
\mbox{ } .  
\eeq

Then, composing, the free-free problem's solution is 
\beq
\rho_2 = \mbox{const}\rho_1 + \mbox{Const} \mbox{ } , \mbox{ } \Phi \mbox{ fixed .}
\eeq
The HO-free problem's solution is  
\beq
\rho_1 = \sqrt{\frac{\fE_1}{H_1}}\mbox{sinh}
\left(
\sqrt{\frac{\mu_2}{\mu_1}\frac{H_1}{\fE_2 + k_2}\{\rho_2 - \mbox{ const}\}}
\right) 
\mbox{ } , \mbox{ } \Phi \mbox{ fixed .}
\eeq
The special multi-HO problem's solution is  
\beq
\rho_2 = \frac{D^2\{\sqrt{H_1}\rho_1 + \sqrt{\fE_1 + H_1\rho_1^2}\}^{2\xi} + \fE_2}
              {2\sqrt{H_2}D\{2\sqrt{H_1}\rho_1 + \sqrt{\fE_1 + H_1\rho_1^2}\}^{\xi}} 
\mbox{ } , \mbox{ } \Phi \mbox{ fixed ,}
\eeq
where $\xi$ is the constant $\sqrt{{\mu_1H_2}/{\mu_2H_1}}$ (i.e. the frequency ratio) 
and $D$ is a constant of integration.

The attractive Newton--Coulomb-free problem's solution is  
\beq
\rho_2 = \frac{\fE_2 + k_2}{\fE_1}\sqrt{\frac{2\mu_1}{\mu_2}}
\left\{
\frac{\sqrt{\rho_1\{\fE_1\rho_1 + N_1\}}}{2} - \frac{N_1}{\sqrt{\fE_1}}\mbox{ln}
\left(
\frac{N_1}{2\fE_1} + \sqrt{\fE}\rho + \sqrt{\rho_1\{\fE_1\rho_1 + N_1\}}
\right)
\right\} \mbox{ } , \mbox{ } \Phi \mbox{ fixed .}  
\eeq

These all behave as expected.  
For example, in the free--free case, the 3 particles mark out a uniformly  
growing triangle (corresponding to their free motion).   
Or, in the special 2 HO case, the 3 partices mark out a boundedly large oscillating triangle, 
which oscillations repeat themselves or not depending in the usual way on whether the period ratio 
is rational or irrational.    

\mbox{ }

\noindent{\bf Nonzero relative angular momentum}

\mbox{ }

\noindent Let $L_i(\fU(\rho_i)) = L_i(k_i\rho_i^{\alpha}) \equiv L_i^{(\alpha)}$.  
I introduce the constants $D_i = 2\{\fE_i + k_i\}/J$ and the new variables $x_i = \rho_i^2$ and 
$X_i = 2\mu_i\{\fE_i + k_i\}x_i/J^2$.   
Then, dropping the $i's$ on the $K_i^{(\alpha)}$, $L_i^{(\alpha)}$, $\mu_i$, $\fE_i$, $k_i$, $n_i$, 
$H_i$, $D_i$, $X_i$ and $x_i$, the requisite $K^{(\alpha)}$ and $L^{(\alpha)}$ are as follows.   
\beq
L^{(0)} = \mbox{arccos}(1/\sqrt{X})
\mbox{ } , \mbox{ }
K^{(0)} = \sqrt{X_i - 1}/D \mbox{ } .
\eeq
\beq
L^{(-1)} = \mbox{arccos}\left(\frac{\mbox{cos}\psi - e}{1 - e\mbox{cos}\psi}\right)
\mbox{ } , \mbox{ }
K^{(-1)} = \psi - e\mbox{sin}\psi \mbox{ } ,
\eeq
where $e$ is the eccentricity and $\psi$ is the {\it eccentric anomaly} defined such that $\rho = a(1 - e\mbox{cos}\psi)$ for 
$a$ the semi-major axis length of the ellipse. 
\beq
L^{(2)} = \frac{1}{2}\mbox{arccos}
\left(
\frac{\frac{J^2}{\mu\sfE x} - 1}{\sqrt{1 - \frac{2HJ^2}{\mu\sfE^2}}}  
\right)
\mbox{ } , \mbox{ }
K^{(2)} = - \frac{1}{2}\sqrt{\frac{\mu}{2H}}\mbox{arccos}
\left(
\frac{2Hx - \fE}{\fE\sqrt{1 - \frac{2HJ^2}{\mu\sfE^2}}}
\right) 
\mbox{ } .
\eeq
Then, composing, one has the following ERPM solutions (each consists of two relations as 
it is a curve in the reduced configuration space which has 3 dimensions).  

The free--free problem's solution is 
\beq
\Phi - \bar{\Phi} = - \sum_{i = 1}^2 \mbox{arccos}(1/\sqrt{X_i}) \mbox{ } , \mbox{ }
\sqrt{h/D} = \sqrt{X_1 - 1} - \sqrt{X_2 - 1} \mbox{ } , 
\eeq
where $\bar{\Phi}$ is a constant of integration.  
As expected for free motions, these orbits are linear, and in terms of the particle position 
coordinates, again a uniformly expanding triangle is described.

The attractive Newton--Coulomb-free problem's solution is 
\beq
\Phi - \bar{\Phi} = - \mbox{arccos}
\left(
\frac{\mbox{cos}\psi -e}{1 - e\mbox{cos}\psi} - \mbox{arccos}(1/\sqrt{x_2})
\right)
\mbox{ } , \mbox{ }
h = \psi - e\mbox{sin}\psi - \sqrt{x_2 - 1} \mbox{ } .  
\eeq
As a simple particular case, note that this includes a circular orbit  
(that for which the second equation and the second term of the first equation are trivial).  
The HO-free problem's solution is 
\beq
\Phi - \bar{\Phi} = \frac{1}{2}\mbox{arccos}
\left(
\frac{\frac{J^2}{\mu_1\sfE_1x_1} - 1}{\sqrt{1 - \frac{2H_1J^2}{\mu_1\sfE_1^2}}}  
\right)
+ \sqrt{X_2 - 1}
\mbox{ } , \mbox{ }
h = - \frac{1}{2}\sqrt{\frac{\mu_1}{2H_1}}\mbox{arccos}
\left(
\frac{2H_1x_1 - \fE_1}{\fE_1\sqrt{1 - \frac{2H_1J^2}{\mu_1\sfE_1^2}}}
\right) 
+ \mbox{arccos}(1/\sqrt{X_2}) \mbox{ } . 
\eeq
Finally, the special multiple HO problem's solution is 
$$
2\{\Phi - \bar{\Phi}\} = \mbox{arccos}
\left(
\frac{\frac{J^2}{\mu_1\sfE_1x_1} - 1}{\sqrt{1 - \frac{2H_1J^2}{\mu_1\sfE_1^2}}}   
\right)
+ \mbox{arccos}
\left(
\frac{\frac{J^2}{\mu_2\sfE_2x_2} - 1}{\sqrt{1 - \frac{2H_2J^2}{\mu_2\sfE_2^2}}} 
\right)
\mbox{ } , \mbox{ }
$$
\beq
2\sqrt{2}h = \frac{\mu_2}{H_2} \mbox{arccos}
\left(
\frac{2H_2x_2 - \sfE_2}{\fE_2\sqrt{1 - \frac{2H_2J^2}{\mu_2\sfE_2^2}}}
\right) 
- 
\frac{\mu_1}{H_1} \mbox{arccos}
\left(
\frac{2H_1x_1 - \sfE_1}{\fE_1\sqrt{1 - \frac{2H_1J^2}{\mu_1\sfE_1^2}}}
\right) \mbox{ } .  
\eeq
When envisaged in terms of the triangle marked out by the three particle positions, these solutions 
behave as expected as regards the triangle being ever-expanding or oscillatory in size.

As regards the generality of these example, I stress that they all still {\sl separately conserve} 
subsystem angular momenta.  
Extension of the study to $\Phi$-dependent potentials is required in order for the analysis to 
include the significant effect of angular momentum exchange between the two subsystems (see Sec 6 and 
\cite{AngleDep}).    
Issues in the toy modelling of quantum cosmology further motivate 
such a study (see the Conclusion).


\section{Full reduction for 3-particle 2D SRPM}

\subsection{The 2D SRPM}

Barbour's SRPM follows from the Jacobi action 
\beq
\fI_{\sJacobi}[\q_I, \dot{\q}_I, \dot{\a}, \dot{\b}, \dot{\zeta}] = 
\int\d\lambda\fL_{\sJacobi}(\q_I, \dot{\q}_I, \dot{\a}, \dot{\b}, \dot{\zeta}) = 
2\int\d\lambda\sqrt{\fT^{\$}\{\fU + \fE\}}  
\label{SRPMAc}
\mbox{ } , 
\eeq
which I have somewhat reformulated for extra conceptual  clarity.  
$\fT^{\$}$ and $\fU$ are not uniquely defined, because of the transformation 
\beq
\fT^{\$} \longrightarrow \widetilde{\fT^{\$}} = \Omega \fT^{\$} \mbox{ } , \mbox{ } 
\fU + \fE \longrightarrow \widetilde{\fU + \fE} = \{\fU + \fE\}/\Omega \mbox{ } .
\eeq
One perspective then is the {\sl geometrically natural form}, with 0-homogeneous\foo{In this paper, I use 
`n-homogeneous' as a shorthand for `homogeneous of degree n'.} kinetic term  
\beq
\fT^{\$}(\q_I, \dot{\q}_I, \dot{\a}, \dot{\b}, \dot{\zeta}) = 
\frac{1}{2I} \mbox{ }_{\sbm}||\vec{{\cal S}}\bq||^2
\mbox{ } , 
\eeq
where
\beq
\vec{{\cal S}}\q_I = \dot{\q}_{I} - \dot{\a} - \dot{\b} \cr \q_I + \dot{\zeta}\q_{I}
\eeq
is the arbitrary Sim(N, d)-frame corrected velocity for $\zeta$ a dilational auxiliary,\foo{This 
is related to \cite{B03}'s dilational auxiliary $c$ by $\zeta = \mbox{ln}c$.}  
and 0-homogeneous potential term 
\beq
\fV = \fV(\|\q_I - \q_J\|) \mbox{ } .  
\eeq 
This 0-homogeneity perspective is clearly a very natural one, and turns out to be enlightening 
(see below); it is also a new perspective, Barbour's original formulation \cite{B03} being in terms of  
2-homogeneous kinetic terms (multiply the above form by $\Omega = I$) and --2-homogeneous potential terms 
(divide the above form by $\Omega = I$).  
The homogeneity of $\fU + \fE$ is also a condition enforced by requiring the theory to be consistent 
(see below).  
N.B. $I$ turns out to be a conserved quantity in SRPM.  
Thus, using $I$ in the kinetic term is not absurd, nor is whichever potential homogeneity overwhelmingly 
restrictive because of $I$ being available to construct potentials well capable of mimicking standard 
potentials, such as linear combinations of distinct-power-law potentials, over extensive regimes.
E.g. (in the geometrically natural form)  $\fV = K\|\q_1 - \q_2\|^2/I$ behaves much like the 
here-forbidden HO potential $\fV = k\|\q_1 - \q_2\|^2$, or $\fV = K\sqrt{I}/\|\q_1 - \q_2\|$ behaves 
much like the here-forbidden Newton--Coulomb potential $\fV = k/\|\q_1 - \q_2\|$.

I use the Lagrangian form of the constraints which follow from this action (Appendix B.1) to 
eliminate the auxiliaries $\dot{a}$, $\dot{b}$ and $\dot{\zeta}$ in Appendix B.2.
Then in terms of the {\it simple ratio shape variable} 
\beq
{\cal R} =  \sqrt{\frac{I_1}{I_2}} = \sqrt{\frac{\mu_1}{\mu_2}}\frac{\rho_1}{\rho_2} 
\mbox{ } , 
\label{SRSV}
\eeq 
the action is 
\beq
\fI_{\mbox{\scriptsize Jacobi}}[ {{\mbox{{\boldmath${\cal S}$}}}}, \dot{{\mbox{{\boldmath${\cal S}$}}}}] 
= \int\d\lambda\fL_{\mbox{\scriptsize Jacobi}}(\dot{{\mbox{{\boldmath${\cal S}$}}}},  
\dot{{\mbox{{\boldmath${\cal S}$}}}}) = 2\int\d\lambda\sqrt{\fT^{\$}\{\fU + \fE\}} = 
2\int\d\lambda\sqrt{\frac{1}{2}\mbox{ }_{{\mbox{\scriptsize{\boldmath${\cal M}$}}}}
||\dot{{\mbox{{\boldmath${\cal S}$}}}}||^2\{\fU + \fE\}}
\label{redSRPM}
\eeq
for $\dot{{\mbox{{\boldmath${\cal S}$}}}}$ the shape coordinates $\{{\cal R}, \Phi\}$ with respect to 
which $\mbox{\boldmath${\cal M}$}$ has components $\{1 + {\cal R}^2\}^{-2}\mbox{diag}(1, {\cal R}^2)$.  
Or, explicitly,
\beq
\fI_{\mbox{\scriptsize Jacobi}}[{\cal R}, \dot{{\cal R}}, \Phi, \dot{\Phi}] = 
2\int \d\lambda \sqrt{   \frac{1}{2}\frac{            \{\dot{{\cal R}}^2 + {\cal R}^2\dot{\Phi}^2\}    }
                                                 {    \{1 + {\cal R}^2\}^2    }        
\{{\fU}({\cal R}, \Phi)   + \fE\} } \mbox{ } . 
\eeq

Note that one really has a family of conformal geometries rather than just the above.
These are clearly conformally flat, and the distinguished flat representative is 
\beq
\fI_{\mbox{\scriptsize Jacobi}}[{\cal R}, \dot{\cal R}, \Phi, \dot{\Phi}] = 
2\int\d\lambda \sqrt{    {\cal T}\{{\cal U} + {\cal E}  \}  } = 
2\int\d\lambda \sqrt{    \frac{\{\dot{\cal R}^2 + {\cal R}^2\dot{\Phi}^2\}}{2}
                         \frac{  {\fU}({\cal R}, \Phi) + \fE  }{ \{1 + {\cal R}^2\}^2  } } 
\mbox{ } .  
\eeq
Thus we can represent the motion as lying within ordinary unit-mass 2D mechanics (albeit
for some fairly unusual potentials inherited from scale invariance).

Also note that the line element corresponding to ${\mbox{{\boldmath${\cal M}$}}}$, 
\beq
\d\sigma^2 = \frac{\d {\cal R}^2 + {\cal R}^2\d\Phi^2}{\{1 + {\cal R}^2\}^2}
\label{strike}
\eeq
is the natural Fubini--Study metric \cite{FS} on $\mathbb{CP}^{N - 2}$, $N = 3$ and constant curvature 4, 
with corresponding line element 
\beq
\d\sigma^2 = \frac{\{1 + ||{\mbox{\boldmath${\cal Z}$}}||_{\sC}^2\}
||\d{\mbox{\boldmath${\cal Z}$}}||_{\sC}^2 - 
|({\mbox{\boldmath${\cal Z}$}},\d{\mbox{\boldmath${\cal Z}$}})_{\sC}|^2}
{\{1 + ||{\mbox{\boldmath${\cal Z}$}}||_{\sC}^2\}^2} \mbox{ } ,
\eeq
for $({\mbox{\boldmath${\cal W}$}}, {\mbox{\boldmath${\cal Z}$}})_{\sC} = 
\sum_{\bar{a} = 1}^{N -2}\overline{{\cal W}}_{\bar{a}}{\cal Z}_{\bar{a}}$, 
$|| \mbox{ } ||_{\sC}$ the corresponding norm, $\overline{\cal Z}_{\bar{a}}$ the complex conjugate of 
${\cal Z}_{\bar{a}}$ and $|\mbox{ }|$ the complex modulus, as may be verified by setting 
${\cal Z}_{\bar{a}} = {\cal R}_{\bar{a}}e^{i\Phi_{\bar{a}}}$.\foo{Fubini-Study
metrics of this form are available for any $N$ for ratios and relative angles paired 
together as complex coordinates on $\mathbb{CP}^{N - 2}$, which is promising as regards the arbitrary-$N$ 
2D case, as $\mathbb{CP}^{N - 2}$ is 
the reduced configuration space for scale, translation and rotation free shapes in 2D \cite{Shape}.} 
In the present, special one-ratio one-relative-angle case, this simplifies to the form  
\beq
\d\sigma^2 = \frac{|\d{\cal Z}|^2}{\{1 + |{\cal Z}|^2\}^2} \mbox{ } .  
\eeq

An important result for Sec 5 and 7 is that (\ref{strike}) is also the line element of $\mathbb{S}^2$ 
with constant curvature 1/2, corresponding to a stereographic plane-polar representation of the sphere, 
which may thus be recast in standard spherical coordinates, 
\beq
\d\sigma^2 = \frac{1}{4}\{\d\Theta^2 + \mbox{sin}^2\Theta \d\Phi^2\} \mbox{ } ,
\eeq
by the transformation ${\cal R} = \mbox{tan}\frac{\Theta}{2}$.  
Another useful trick is that using not ${\cal R}$ but the upside-down ratio variable 
${\cal W} = \sqrt{I_2/I_1} = 1/{\cal R}$ gives the same geometry but generally maps the original 
potential to a different function.

The above coordinates all admit their standard ranges of validity.   
The undefinedness of the polar angle at the origin ${\cal R} = 0$ corresponds to 
$\rho_2/\rho_1$ blowup i.e. collision of particles 2, 3.  
${\cal R} = \infty$ corresponds to $\rho_1/\rho_2$ blowup, i.e. collinearity of particle 1 and the centre 
of mass of particles 2, 3.     
%
%
That these are not places where serious problems occur can be seen by noting by cyclic permutation of 
particle labels that there are two other places where each such occurs, 
which the above coordinate patch covers and sees no pathological behaviour thereat.  
%
%
This is not however advantageous in the below study.  
The linking conformal factor to get to the flat form is $I^2$. 
While conformal transformations are well-known in mechanics and GR of being capable of excluding 
physically-relevant regions, the situation here is as follows. 
With the $\rho_i$ already being defined as radii and hence non-negative, their ratio is nonnegative.
Thus the conformal factor cannot be zero.  
It can be infinity: at ${\cal R} = \infty$ (but there's nothing `beyond' that is excluded by the 
conformal transformation).  
Elsewhere it is smooth.
Thus making this conformal transformation does not amount to throwing away any regions.

I provide the variation of the fully reduced SRPM action in Appendix B.3.

\section{Simplifications to SRPM equations of motion for angle-free potentials}

\subsection{The `central force' simplification} 

In the case of 
${\cal U} = {\cal U}({\cal R})$ alone, the $\Phi$ Euler--Lagrange equation (\ref{PhiEL}) simplifies to  
\beq
P_{\Phi} = {\cal R}^2\Phi^{\prime} = {\cal J} \mbox{ , constant ,}
\label{ha}
\eeq  
which is the mathematical analogue of conservation of angular momentum in $\{{\cal R}, \Phi\}$ space.  
Then, substituting this into the problem's other first integral (\ref{OI}), I obtain  
\beq
\frac{1}{2}
\left\{
{\cal R}^{\prime 2} + \frac{{\cal J}^2}{{\cal R}^{2}}
\right\} 
+ {\cal V}({\cal R}, \Phi) = {\cal E}({\cal R}) \mbox{ } 
\eeq
as the remaining equation to be solved.  
%
%
The quadrature for the shape of the orbit is thus 
\beq
\Phi - \bar{\Phi} = 
\int \frac{{\cal J}\d {\cal R}}{{\cal R}\sqrt{2\{{\cal U}({\cal R}) + {\cal E}({\cal R})\}{\cal R}^2 - {\cal J}^2}} = 
- {\cal J}\int\frac{\d{\cal W}}{\sqrt{2\{{\cal U}({\cal W})+{\cal E}({\cal W})\} - {\cal J}^2{\cal W}^2}} 
\mbox{ } , 
\eeq 
${\cal W}  = 1/{\cal R}$ being a useful change of variables     
%
%
because the potentials considered below are simpler in terms of ${\cal W}$.

\subsection{Some physically interesting potentials}

Let's consider a fairly but not entirely general class among the $\Phi$-free 0-homogeneous 
(negative) potential contributions: $\fU = -k\left\{\frac{\rho_1}{\sqrt{I}}\right\}^{\alpha}$.  
In the planar representation, these become  
\beq
{\cal U}({\cal R}) = Q\frac{{\cal R}^{\alpha}}{\{1 + {\cal R}^2\}^{\alpha/2 + 2}  } \mbox{ } , \mbox{ }
Q \equiv \frac{k}{\mu_1^{\alpha/2}} \mbox{ } .  
\label{genpot} 
\eeq
Also, in the spherical representation, they are 
\beq
\fU(\Theta) = Q\mbox{sin}^{\alpha}\mbox{$\frac{\Theta}{2}$} \mbox{ } .  
\eeq 
E.g. for the single HO-like potential, 
${\cal U} = -H_{1}{\cal R}^2/\{1 + {\cal R}^2\}^3$ and $\fU = H_{1}\frac{\mbox{cos}\Theta - 1}{2}$.  
E.g. for the single Newton--Coulomb-like potential, 
${\cal U} = \frac{N_{1}}{{\cal R}\{1 + {\cal R}^2\}^{3/2}}$ and $\fU = N_{1}\mbox{cosech}\frac{\Theta}{2}$.  
[A complementary class of potential contributions that I make rather less use of are 
$\fV = -k\left\{\frac{\rho_2}{\sqrt{I}}\right\}^{\alpha}$.  For these, 
\beq
{\cal V}({\cal R}) = -\frac{Q}{\{1 + {\cal R}^2\}^{\alpha/2 + 2}} \mbox{ } , \mbox{ } 
\fV(\Theta) = Q\mbox{cos}^{\alpha}\mbox{$\frac{\Theta}{2}$} \mbox{ } .] 
\eeq

The asymptotic behaviour of the potentials is $\fE + \fU({\cal R}) \approx \fE + Q{\cal R}^{\alpha}$ 
for ${\cal R}$ small (which is a standard problem), ${\cal V}({\cal R}) \approx \{\fE + Q\}/{\cal R}^4$ 
for ${\cal R}$ large, i.e. a shared standard problem.

\subsection{The significant potential-like quantities}

Define ${\cal V}_{\eff} \equiv {\cal V} + \frac{{\cal J}^2}{{\cal R}^2} - {\cal E}$, which 
is the potential quantity that is significant for motion in time and at the quantum level.  
Also define ${\cal U}_{\orb} \equiv - {\cal R}^4{\cal V}_{\eff}$, which is the potential quantity 
that is significant as regards the shapes of the classical orbits.  
The whole-universe aspect of our modelling means that we wish to 
study these with fixed $\fE$, and free parameter ${\cal J}$, which is somewhat unusual.

For the physically interesting potentials considered in this paper and their small and large ${\cal R}$ 
limits, I sketch ${\cal V}_{\eff}$ and ${\cal U}_{\orb}$ in Fig 2, placing emphasis on the qualitative 
differences between the various cases. 
The shaded lines on the ${\cal U}_{\orb}$ sketches indicate the classically forbidden regions 
and hence whether the orbits are bounded from above and/or below certain values of ${\cal R}$.  

\mbox{ }

\noindent[{\bf Caption for figure 2.} ${\cal V}_{\eff}$ and ${\cal U}_{\orb}$ for

\noindent i) the large ${\cal R}$ asymptotic solution.  ${\cal V}_{\eff}$ 
exhibits a finite potential barrier and the orbits are bounded from above.    

\noindent ii) The small ${\cal R}$ asymptotic solution 
to the $\fE = 0$ attractive Newton--Coulomb problem is the usual Newton--Coulomb problem. 
${\cal V}_{\eff}$ has a potential trough with an infinite centrifugal potential barrier inside it, 
and the orbits are bounded from below.

\noindent iii)   The $\fE = 0$ attractive Newton--Coulomb SRPM problem.  
In contrast with the usual Newton--Coulomb problem, ${\cal V}$ dips down to $-\infty$ as 
${\cal R} \longrightarrow \infty$, and the orbits are bounded from above as well as from below.  

\noindent iv) The small ${\cal R}$ asymptotic solution for a fixed $\fE > 0$ 
HO problem is the usual radial HO problem.  
${\cal V}_{\eff}$ is an infinite well formed by the HO's parabolic potential on the outside 
and the infinite centrifugal barrier on the inside, and the orbits are bounded from below.   

\noindent v) A fixed $\fE > 0$ HO SRPM problem.  
In contrast with the usual radial HO problem, the potential tends to 0 rather than $+\infty$ as 
${\cal R} \longrightarrow \infty$, and the orbits are bounded from both above and below.  
Also, its Newton--Coulomb-like ${\cal V}_{\eff}$ means that it, unlike the usual radial HO, 
is `ionizable' -- sufficiently high energy quantum states can now escape from the well.]

\subsection{Some simple exact solutions and exact asymptotic solutions}  

\noindent{\bf Zero relative angular momentum}

\mbox{ }

\noindent There being no centrifugal potential, this coincides with the simpler case of motion 
in a 1-d potential.  
%
%

\mbox{ } 

\noindent{\bf Nonzero relative angular momentum}

\mbox{ } 

\noindent With the potential (\ref{genpot}), the quadrature for the shape of the orbit take the form 

\noindent
\beq
\Phi-\bar{\Phi} = 
{\cal J}\int\frac{\d {\cal R}\{1 + {\cal R}^2\}^{\frac{\alpha}{4} + 1}}
{{\cal R}\sqrt{2{\cal R}^2\{Q{\cal R}^{\alpha} + \fE\{1+{\cal R}^{\frac{\alpha}{2}}\} - 
{\cal J}^2\{1+{\cal R}^2\}^{\frac{\alpha}{2}+2} }} 
= -{\cal J}\int\frac{\d {\cal W}\{1 + {\cal W}^2\}^{\frac{\alpha}{4} + 1}}
                {\sqrt{2{\cal W}^{4}\{Q+\fE\{1+{\cal W}^2\}^{\frac{\alpha}{2}} - 
{\cal J}^2\{1+{\cal W}^2\}^{\frac{\alpha}{2}+2} }} 
\mbox{ } .
\eeq

Then $\fE = 0$ and $\fU = kI^2/||q_2 - q_3||^4$ (which maps to $Q/{\cal R}^4$) gives the exact SRPM 
orbits 
\beq
{\cal W} = \frac{\beta}{\sqrt{2}}\mbox{sec}(\Phi - \bar{\Phi}) 
\eeq
or
\beq
{\cal R} = \sqrt{2}\gamma\mbox{cos}(\Phi - \bar{\Phi}) 
\mbox{ } , 
\label{exactsoln}
\eeq
for $\beta = \frac{1}{\gamma} = \frac{{\cal J}}{\sqrt{Q}}$ a dimensionless parameter.  
In the first representation, these orbits are a family of straight lines in polar coordinate 
form, corresponding to geodesic motion in the flat representation.  
In the second representation, these orbits are a family of circles of radius $\gamma/2$ and 
centre $(\gamma/2, 0)$ (i.e. they are all tangent to the vertical axis through the origin).   
These orbits play an important role as a limiting case in the below analysis.  
Note also that 
\beq
{\cal R} = \frac{{\cal J}}{\sqrt{2Q}}\mbox{sec}(\Phi - \bar{\Phi}) 
\eeq
is also an exact SRPM orbit (as in this case ${\cal R}$ and ${\cal W}$ are interchangeable in the 
potential as well as in the kinetic term.

Next, as noted above, for ${\cal R}$ large ($>> 1$), const$/{\cal R}^4$ is the approximate form of {\sl any} 
${\cal U} + {\cal E}$.  
Thus (\ref{exactsoln}) is an asymptotic solution, assuming that it is classically allowed 
[in some cases the integral goes complex before one gets to the ${\cal R}$ large regime (confirmed by 
Maple); classically-allowed corresponds to ${\cal U}({\cal R}) {\cal R}^2 - {\cal J}^2 > 0$.   
On the other hand, for ${\cal R}$ small ($<< 1$), (\ref{genpot}) takes the more standard approximate form 
\beq
{\cal U} \approx Q{\cal R}^{\alpha} + \fE \mbox{ } , 
\eeq
for which one can borrow results from standard dynamics literature (see e.g. \cite{Goldstein, Moulton}).  
Then e.g. $\alpha = 2, -1, -2$ are exactly soluble in terms of elementary functions for $\fE \neq 0$, 
while there is a wider range of $\alpha$'s for which the $\fE = 0$ problems are exactly soluble in terms 
of elementary functions.  
The cases of relevance for this paper are $\alpha = 2$, for which the small-${\cal R}$ orbits 
are nested ellipses centred about the origin, and $\alpha = - 1$, for which the orbits are progressively 
less eccentric hyperbolae nested outside a parabola nested outside progressively less eccentric 
ellipses ending with an innermost circle.

An exact SRPM orbit solution illustrating interpolation between small and large {\cal R} behaviours is 
for $\fE = 0$, $\fU = kI^2/||q_2 - q_3||^6$ which maps to ${\cal U} = Q(1 + {\cal R}^2)/{\cal R}^6$.  
This gives a ${\cal V}_{\eff}$ of the same qualitative type as the first ${\cal V}_{\eff}$ curve in 
Fig 2, and a ${\cal U}_{\orb}$ of the same qualitative type as the third ${\cal V}_{\eff}$ curve in Fig 2.  
I can solve for this using form 2 of the quadrature, obtaining  
\beq
{\cal R} = \frac{1}{\beta}\sqrt{\sqrt{1 + 2\beta^2}\mbox{cos}(2\{\Phi - \bar{\Phi}\}) + 1}
\eeq
Now the radicand is bounded to lie between $1 \pm \sqrt{1 + \beta^2}$.  
${\cal R}_{\mbox{\scriptsize max}} = \frac{1}{\lambda}\sqrt{1 + \sqrt{1 + 2\beta^2}}$.  
Whether ${\cal R}$ is allowed to be large or small depends on $\beta$.  
Note that ${\cal R}$ is a radius, so is $\geq 0$.  
That means that there is a critical angle beyond which there is no solution, 
$\Phi - \bar{\Phi} = \frac{1}{2}\mbox{arccos}\left(- \frac{1}{\sqrt{1 + 2\beta^2}}\right)$.  
For $\beta \approx \infty$, corresponding to small ${\cal R}$ like behaviour, ${\cal R} \approx 
\sqrt{\sqrt{2}\mbox{cos}(2\{\Phi - \bar{\Phi}\})/\beta}$, the critical angle is $\pi/4$.  
The small-${\cal R}$ asymptotic behaviour is thus a family of nested tear drops with a common tip 
at the origin and common tangents there that are inclined at $\pi/2$ to each other.    
For $\beta = 0$, corresponding to large ${\cal R}$ like behaviour, ${\cal R} \approx 
\frac{\sqrt{2}}{\beta}\mbox{cos}(\Phi - \bar{\Phi})$, the critical angle is $\pi/2$, 
corresponding to the usual large ${\cal R}$ regime of nested circles centred along a single line 
that touch at the origin.  
The interpolatory behaviour exhibited by the full solution is a nested family of tear drops with a 
common tip for which the two tangents at the tip become more obtuse as ${\cal R}$ increases, 
progressing from forming an angle of $\pi/2$ at large $\beta$ toward forming an angle of $\pi$ for small 
$\beta$.  
This corresponds to the two tangents having finally joined up into a single line, by which 
stage the outermost `tear drops' have been fully smoothed out into the large-${\cal R}$ circles.  
My exact calculations for this potential give that the case with ${\cal R}_{max} = 0.01$ has maximum 
angle 45.002 degrees, 
which suggests that not many orders of magnitude need be considered in numerically investigating the 
interpolating behaviour in the cases of the more physically interesting potentials considered below.


$\alpha = 0$ yields another exact solution: the problem of motion in the spherical representation 
is equivalent to the geodesic problem on the sphere and thus solved by great circles, e.g.
\beq
\mbox{cos}(\Phi - \bar{\Phi}) = C\mbox{cot}\Theta = \frac{C}{2{\cal R}}\{1 - {\cal R}^2\}  
\eeq  
are great circles for $C$ constant.
Useful relations between the two representations include that ${\cal R} = 1$ is the equator,   
${\cal R}$ small means well within the northern hemisphere, 
and {\cal R} large means well within the southern hemisphere.
%

\subsection{Numerical investigation}

From Subsec 5.4, a first guess is that numerics should concern `several orders of magnitude in 
${\cal R}$', `centred' about 1, e.g. ${\cal R} = 10^{-2}$ through to $10^2$.
I chose to integrate the first of the above quadratures for the single attractive Newton--Coulomb 
and single HO cases of most interest physically, using Maple's Runge--Kutta solver \cite{Maple}.  
[This method was tested by applying it to the previous section's exact solutions, resulting in agreement 
with the exactly-evaluated shapes of the orbits.]

In the attractive Newton--Coulomb-like SRPM problem that is analogous to the standard parabolic slingshot, 
Instead of being unbounded like the slingshot, I find that the new orbit curves inwards from that 
and then spirals around to a maximum value around which the orbit is well-approximated by the outer piece  
of a large ${\cal R}$ regime circular orbit.   
The maximum value and the amount of spiralling both depend on the value of $\beta$.  
E.g. for $\fE = 0$ and $\beta^2 = 0.1$, it spirals around one revolution on its way out 
to a maximum value of around 4.3${\cal R}$. 
While, as one moves down to $\fE = 0$ and $\beta^2 = 0.01$, it spirals slightly over 2 revolutions 
on its way out to a maximum of around 14${\cal R}$.

In the HO-like SRPM problem, for small ${\cal R}$ the motion begins well-approximated by the elliptical 
orbit centred at the origin of the corresponding standard motion, but then curves outwards from this 
and spirals to a maximum value around which the orbit is well-approximated by the outer piece of a large 
${\cal R}$ regime circular orbit, before returning along a similar trajectory to the elliptical shape 
of the small ${\cal R}$ regime.  
Again, the smaller $\beta$ is in the SRPM model, the more spiralling there is before the outflung 
object returns, and the maximum value attained is larger.  
To get this case to work, one needs to prescribe a value for $\fE/Q$ in addition to prescribing 
$\beta$.  
For example, for $\fE/Q = 100$ and $\beta^2$ = 0.1, it spirals around 1 revolution before reaching 
a maximum value of around 44${\cal R}$, while if $\beta^2$ is decreased to 0.01, it spirals 
over 2 revolutions before reaching a maximum value of around 140${\cal R}$.

\section{Posing the $\Phi$-dependent SRPM triple HO and triple Newton--Coulomb problems}

As regards interesting $\Phi$-dependent potentials, the triple HO-like potential 
$\fV = h_{1}\{\q_2 - \q_3\}^2 +$ cycles maps to 
\beq
{\cal V} = \frac{\mu_2 H_1{\cal R}^2 + G\sqrt{\mu_1\mu_2}{\cal R}\mbox{cos}\Phi + \mu_1H_2}
{2\mu_1\mu_2\{1 + {\cal R}^2\}^3} 
\eeq
in the planar representation.
Also, in the spherical representation, it is
\beq
\fV = A + B\mbox{cos}\Theta + C\mbox{sin}\Theta\mbox{cos}\Phi
\eeq
for
\beq
A = \frac{1}{4}
\left\{
\frac{H_1}{\mu_1} + \frac{H_2}{\mu_2} 
\right\}
\mbox{ } , \mbox{ }
B = \frac{1}{4}
\left\{
\frac{H_2}{\mu_2} - \frac{H_1}{\mu_1} 
\right\}
\mbox{ } , \mbox{ } 
C = \frac{G}{4\sqrt{\mu_1\mu_2}}
\mbox{ } . 
\eeq

The triple Newton--Coulomb-like potential $\fV = -n_{23}/||\q_2 - \q_3|| +$ cycles maps to 
\beq
{\cal U} = \frac{\mu_1}{\{1 + {\cal R}^2\}^{3/2}}
\left\{
\frac{N_{1}}{{\cal R}} +
\frac{N_{2}}{\sqrt{C^2 + 2UC{\cal R}\mbox{cos}\Phi + U^2{\cal R}^2}}                           +
\frac{N_{3}}{\sqrt{C^2 + 2VC{\cal R}\mbox{cos}\Phi + V^2{\cal R}^2}}                           
\right\}
\eeq
in the planar representation.
In the spherical representation, one has 
\beq
\fU =  \sqrt{\mu_1}\mbox{cosec}\mbox{$\frac{\Theta}{2}$}
\left\{ 
N_{1} + N_{3}
\left\{
C^2\mbox{cot}^2\mbox{$\frac{\Theta}{2}$} + 2UC\mbox{cot}\mbox{$\frac{\Theta}{2}$}\mbox{cos}\Phi + U^2 
\right\}^{-1/2}                           
+ N_{2}
\left\{
C^2\mbox{cot}^2\mbox{$\frac{\Theta}{2}$} + 2VC\mbox{cot}\mbox{$\frac{\Theta}{2}$}\mbox{cos}\Phi + V^2 
\right\}^{-1/2}                           
\right\}
\mbox{ } .  
\eeq
One would solve each of these with $\fE$ term (add $\fE/\{1 + {\cal R}^2\}^2$ 
to the planar representation's ${\cal U}$ or $\fE$ to the spherical representation's $\fV$).

The asymptotic regimes are as for the single potential cases: an energy-like constant plus a standard 
potential for ${\cal R}$ small, and const$/{\cal R}^4$ for ${\cal R}$ large, so the present paper's 
methods look to be a solid base for this more complicated investigation, which will appear in a further 
paper \cite{AngleDep}.

%
%
%
%
%
%


\section{Conclusion}

For Euclidean relational particle mechanics (ERPM) and similarity relational particle mechanics (SRPM), 
I have provided full reductions for the `triangle land' \cite{EOT} cases of 3 particles in 2D.

SRPM is a mathematically new problem, characterized by its new class of potentials that arise from scale 
relationalism implying both stringent homogeneity requirements on the class of potentials allowed and 
yet also the existence of a conserved quantity, the moment of inertia, which allows this class to be 
broad enough to contain potentials that mimic the standard potentials of mechanics in the regime 
$1 >>{\cal R}$ (the ratio of the sizes of the two relative Jacobi coordinates).
I have provided a number of exact solutions to ERPM and SRPM as well as investigating the new and 
physically interesting cases of Newton--Coulomb-like and harmonic oscillator-like potentials in SRPM.  
In each case there is contact with standard mathematics, at least for 3 particles in 
2D.\foo{The 3D case is expected to have harder mathematics \cite{Shape, FORD}, with the 3D SRPM quite 
possibly being new in its mathematical form as a dynamical system.}

This mathematics being shared with part of that arising in the absolutist approaches to mechanics, 
I make the philosophically significant comment that there is one sense in which the historical 
pre-eminence of absolutism has not harmed the development of physics:    
absolutism may be regarded as a simple path to discovering dynamical systems which then take on a 
mathematical life of their own and are found to be applicable to many other settings,
while relationalism is a conceptually cleaner but more complicated (and hence historically much later) 
path which nevertheless tends to lead to the {\sl same sort of body of mathematics}.  
[Relationalism is, nevertheless, a conceptually important {\sl foundational alternative} to 
absolutism, and, in the way of well-thought-out reformulations \cite{reform}, it may lead to new insights 
and new results.]

In ERPM, the connection with conventional Newtonian mechanics is clear.  
{\sl ERPM is the recovery of the mathematics of a portion of Newtonian mechanics, now on a relational footing.}

{\sl SRPM is new, however, through scale relationalism not being part of the conventional mechanics.}  
Nevertheless, there are some connections with the mathematics of a portion of Newtonian mechanics: 
for this paper's 2D examples, each case becomes a different well-known dynamics in the $1 >>{\cal R}$ 
regime. 
Also, for all allowed potentials, there is a single, shared well-known dynamics in the 
$1 << {\cal R}$ regime.  
The novelty of SPRM is in the {\sl transition} from standard small-scale behaviour to this 
universal large-scale behaviour (which is in general quite distinct from the small-scale behaviour).  
This serves as an intriguing suggestion of how a symmetry principle is capable of reproducing 
standard physics at smaller scales while diverging significantly at larger scales.  
This is due to the `wider matter distribution in the universe' (here particle 1) affecting the 
physics of other subsystems (here particles 2 and 3) on large enough scales, 
which is interestingly `Machian' (in another sense of the word from that used in the Introduction 
\cite{buckets}).  
This may have some capacity to account for deviations from standard physics at larger scales without 
having to invoke (as many instances of) dark matter, e.g. explaining (at least at a nonrelativistic level) 
the rotation curves of galaxies without incurring unacceptably large deviations in solar system physics.  
But clearly one needs to advance from the present paper's setting to many particles in 3D, and consider 
which of a realistic distribution of distant masses can be robustly neglected or agglomerated,  
before reliable quantitative calculations of the consequences of putative scale invariance for galaxies and the solar system 
can be done.  
In this direction, I note that there are indications that scale relationalism and rotational relationalism 
do not interfere with each other \cite{06II}.    
Rotational relationalism being what greatly complicates 3D treatment, some indication of whether scale 
relationalism has important physical consequences for many particles in 3D may be found by considering 
scale and translation (but not rotation) relational particle mechanics.  
That is work in progress, started in \cite{FORD}.

Other applications of the present paper to the investigation of conceptual issues in theoretical 
physics are 1) its quantum counterpart \cite{07II} for its own sake as an example of global and  
operator-ordering issues with quantization.  
The variables and geometrizations introduced in the present paper are important in this study.  
2) The ERPM scale variable (App C) can be rearranged to be an internal time variable at the classical level 
\cite{06II, SemiclI}, thus furbishing a toy model for further investigation of the hidden internal time 
approach to the problem of time in quantum gravity \cite{Kuchar92, Isham93}.  
3) This quantum counterpart of this paper can furthermore be used as a toy model of the semiclassical 
approach (see e.g. \cite{Kieferbook}) to the problem of time in quantum gravity and quantum cosmology 
\cite{SemiclIII}, in which role it enjoys the advantage \cite{SemiclIII} of sharing more relevant 
features with GR than the 1-d relational particle models previously studied in this way \cite{SemiclI}.  
In particular, it serves to exemplify the further complications that arise when the Jacobi--Synge 
metric depends on the light d.o.f. freedom as well as the heavy ones and thus cannot be pulled out 
of the light d.o.f. expectation values, and when the Laplacian ordering is employed rather than the 
simple `momenta to the right' one.   
Nonseparability, as afforded e.g. by the extension of the present paper to $\Phi$-dependent potentials 
(as posed in Sec 6 and investigated in \cite{AngleDep}) is an important aspect of this study.  
4) This extension also amounts to being a model of relative angular momentum exchange dynamics within an 
overall-timeless model.  
This is interesting at both the classical and the quantum level as a dynamically nontrivial model 
for the timeless records theory approach\foo{See \cite{B94II, EOT, PW, Rec, Records} for various 
records theory approaches; \cite{QCosEv} also argue that at least some of these approaches are of 
interest in quantum cosmology.}  to the problem of time, in which one is to study how much 
dynamical or historical information can be extracted solely from the correlations between 
subconfigurations.  
At the quantum level it would furthermore be interesting as an arena for investigating 
the conditional probabilities interpretation \cite{PW}.  
The value to records theory of this kind of model increases if one passes from 3 to $N$ particles, 
as I do classically in \cite{FORD} in the 2D case, as then various particle clusters serve as    
`information-containing localized subconfigurations of a single instant', which is what records are.   
This $N$-body study in 2D is likely to benefit from some sophisticated tests for the significance of 
patterns at the classical level.\foo{In \cite{Shape} and references therein, whether collinearities are 
statistically significant is studied in the context of 2D shape spaces.}
5) The $N$-body study in 2D could also be used as an analogue of the investigation of 
whether microsuperspace dynamics lies stably within minisuperspace dynamics \cite{KR89}, 
by looking e.g. at whether the $N$ = 3 model lies stably within the $N$ = 4 one.    

\mbox{ }

\noindent{\bf Acknowledgments}

\mbox{ }

\noindent I thank Dr. Julian Barbour for the idea of working on relational particle models with and 
without scale, Professor Don Page for discussions about how to study triangles, spheres and projective 
spaces, Dr Terry Rudolph for discussions and Mr Simeon Bird for directing me to the Maple Runge--Kutta 
solver.  I also thank Peterhouse for funding in 2006--2007.

\mbox{ }

\noindent{\bf{\large Appendix A. Manipulating the ERPM action into fully reduced form}}

\mbox{ }

\noindent{\bf A.1 Variation of the ERPM action (\ref{A1})}

\mbox{ }

\noindent The momenta are
\beq
\p^I = \delta^{IJ}m_I\vec{{\cal E}}{\q}_{J}/\dot{\mbox{I}} 
\eeq
for $\dot{\mbox{I}} = \sqrt{\frac{\sfT}{\sfU + \sfE}}$ the `emergent lapse' (emergent time elapsed).    
They satisfy the primary constraint
\beq
\ttH(\q_I,\p^I) = \mbox{ }_{\sbn}\|\bp\|^2/2 + \fV(||\br||) = \fE \mbox{ } .
\eeq
Free end point\foo{Free end point variation \cite{ABFO, ABFKO} with respect to a variable $g$ means that 
\beq
0 = \delta \fI = \int_{\lambda_i}^{\lambda_j}\d\lambda
\left\{
\frac{\pa\fL}{\pa g} - \frac{\d}{\d\lambda}\frac{\pa\fL}{\pa \dot{g}}
\right\}\delta g
+ 
\left\{
\frac{\pa\fL}{\pa\dot{g}}\delta g
\right\}_{\lambda_i}^{\lambda_f}
\eeq
leads to three equations rather than the usual one, because the set of varied curves 
under consideration is more general than usual.    
If, as always in this paper, the $g$ is cyclic, these are ${\pa\fL}/{\pa \dot{g}} =$ constant, 
$\left.{\pa\fL}/{\pa \dot{g}}\right|_{\lambda_i} = 0 = \left.{\pa\fL}/{\pa \dot{g}}\right|_{\lambda_f}$
so, overall, ${\pa\fL}/{\pa \dot{g}} = 0$.  
If $g$ is a cyclic gauge auxiliary, free end point variation is part of the embodiment of the gauge principle.  
Note that while this is unusual, 1) it does agree with what happens if one supplants the $\dot{g}$'s 
in the equations by a multiplier $h$: the multiplier equation is then ${\pa\fL}/{\pa h} = 0$ with common 
rather than free endpoint variation sufficing. 
2) The latter but not the former spoils the manifest reparametrization invariance of the action.  
It is because of this that I choose the former, and hence my precise choice of action \ref{A1} rather 
than Barbour--Bertotti's original action.   
See \cite{FEP} for further details.} 
variation with respect to $\a$ gives  
\beq
\underline{\ttP}(\q_I, \p^I) \equiv \sum_{I = 1}^N \p_I = 0 \mbox{ } , 
\label{ZM}
\eeq
i.e. that the system is constrained to have zero total momentum.   
Free end point variation with respect to $\b$ gives  
\beq
\underline{\ttL}(\q_I, \p^I) = \sum_{I = 1}^N \q_I \cr \p^I = 0
\label{ZAM}
\eeq
i.e. that the system is constrained to have zero total angular momentum. 
[In 2D, free end point variation with respect to $b \mbox{ } ( = b_3)$ produces a single component zero angular 
momentum constraint $\ttL(\q_I, \p^I) = \sum_{I = 1}^N \{q_{I1}p^{I2} - q_{I2}p^{I1}\} = 0$ while in 1D 
there is neither any auxiliary rotational variable with respect to which to vary nor a notion of angular 
momentum.]    
The Euler--Lagrange equations obtained from variation with respect to $\q_I$ propagate all these 
constraints.

\mbox{ }

\noindent{\bf A.2 Elimination of the translations and rotations}

\mbox{ } 

\noindent The Lagrangian form of (\ref{ZM}), 
\beq
\sum_{I = 1}^{N} m_I\{\dot{\q}_{I} - \dot{\a} - \dot{\b} \cr \q_{I}\}  = 0
\eeq 
can readily be used to eliminate $\dot{\a}$ from the action (\ref{A1}) by Routhian reduction.  
One obtains thus the action in terms of a still fairly redundant set of $DN(N -1)/2$ relative position 
variables,   
\beq
\fI_{\sJacobi}[\r_{IJ},\dot{\r}_{IJ}, \dot{\b}] = 
\int\d\lambda\fL_{\sJacobi}(\r_{IJ},\dot{\r}_{IJ}, \dot{\b}) =
\int\d\lambda\sqrt{\fT\{\fE + \fU\}}
\eeq
where  
\beq
2\fT(\r_{IJ},\dot{\r}_{IJ}, \dot{\b}) = \mbox{}_{\sbfm}\|\vec{{\cal R}}\br\|^2
\mbox{ } , 
\eeq
for $\bfm$ the relative mass matrix with components $\frac{1}{2M}\{m_Im_J - m_I^2\delta_{IJ}\}$ 
in the $\r_{IJ} \equiv \q_I - \q_J$ coordinate system, and 
\beq
\vec{\cal R}\underline{r}_{IJ} = \dot{\underline{r}}_{IJ} - \dot{\b} \cr \underline{r}_{IJ}
\eeq
the Rot(D)-frame corrected velocities.  
This can be recast diagonally and in close mathematical analogy with the absolute problem with D 
extra coordinates, by passing to D$(N - 1)$ relative Jacobi coordinates extended by whatever rotational 
auxiliaries $\b$ exist in dimension D: 
\beq
\fI_{\sJacobi}[\R_i, \dot{\R}_i, \dot{\b}] = \int\d\lambda\fL_{\sJacobi}(\R_i, \dot{\R}_i, \dot{\b}) = 
2\int\d\lambda\sqrt{\fT\{\fE + \fU\}} \mbox{ } ,
\eeq
with dikinetic energy
\beq
2\fT(\dot{\R}_i, \b) = \mbox{}_{{\mbox{\scriptsize{\boldmath$\mu$}}}}\|\vec{{\cal R}}{\bR}\|^2
\eeq
where the Jacobi relative mass matrix is $\mbox{\boldmath$\mu$} = \mbox{diag}(\mu_i)$, 
and $\fU$ is now of form $\fU(\|R_i\|, \R_i\cdot \R_j)$.  
The momenta are

\noindent
\beq
\P^i = \delta^{ij}\mu_i\vec{{\cal R}}\R_j/\dot{\mbox{I}} \mbox{ } .
\eeq
From these again follows a primary constraint, 
\beq
\ttH(\R_i, \P^i) = \mbox{}_{_{\mbox{\scriptsize{\boldmath$\nu$}}}}\|\bP\|^2/2 + \fU =\fE \mbox{ } , 
\eeq
where ${\mbox{\boldmath$\nu$}}$ is the inverse of ${\mbox{{\boldmath$\mu$}}}$.  
Free end point variation with respect to $\b$ yields the zero angular momentum constraint in Jacobi coordinates  
\beq
\underline{\ttL}(\R_i, \P^i) = \sum_{i = 1}^n \R_i \cr \P^i = 0 \mbox{ } ,  
\label{Lag}
\eeq
in 3D. $\ttL(\R_i, \P^i) = \sum_{i = 1}^n \{R_{i1}P^{i2} - R_{i2}P^{i1}\} = 0$ in 2D, and in 1D 
there is no reduction to perform.

The Lagrangian form of (\ref{Lag}), 
\beq
\sum_i\mu_i\{\R_i \cr \dot{\R_i} - \R_i \cr \{ \dot{\b} \cr \R_i \}\} = 0
\eeq
can then be used to eliminate $\dot{\b}$ from the Jacobi action by Routhian reduction.
This produces the Jacobi action (\ref{ARot}).    

\mbox{ }

\noindent{\bf A.3 Reformulation of ERPM terms of independent relational variables}

\mbox{ }

\noindent For the ($N = 3$ i.e. $n = 2$) case in 2D, I perform the following coordinate transformations.  
\noindent 1) I transform into relative Jacobi bipolar coordinates   
\beq
\uR_i = (\rho_i\mbox{cos}\theta_i, \rho_i\mbox{sin}\theta_i) \mbox{ } ,  
\eeq
for which 
%
%
(\ref{fT}), (\ref{2Deasier}) give a kinetic term depending on $\rho_i$, $\dot{\rho}_i$ and 
$\dot{\theta}_i)$ alone.  
%

\noindent 
2) I then transform into the redefined angular coordinates  
\beq
\Phi = \theta_2 - \theta_1       
\mbox{ } , \mbox{ } 
\Psi = \theta_2 + \theta_1 \mbox{ } ,   
\eeq
the first of which is entirely relational (c.f. Figure 1) while the second contains absolute information.  
Using this transformation is an `isolation of the absolute vestige' technique similar to those used in 
\cite{06I}; it works out as follows.  
The manifestly relational expression discards the variable in which the absolute vestige itself is 
isolated: overall, the $\dot{\theta}_i$ contribute {\sl only} $\dot{\Phi}$ terms.  
Then by basic algebra one is left with  
\beq
2I\fT(\rho_1, \rho_2, \dot{\rho}_1, \dot{\rho}_2, \dot{\Phi}) = 
E^2 + I_1I_2\dot{\Phi}^2 =
IP + I_1I_2 \dot{\Phi}^2 \mbox{ } , 
\eeq
or, in geometrical form, 
\beq
2\fT({\mbox{\boldmath${\cal R}$}}, \dot{\mbox{\boldmath${\cal R}$}}) = 
\mbox{}_{\mbox{\scriptsize{\boldmath${\cal M}$}}}\|\dot{{\cal R}}\|^2 \mbox{ } 
\eeq
for mass matrix ${\mbox{\boldmath${\cal M}$}}({\mbox{\boldmath${\cal R}$}})$ with components  
$\mbox{diag}(\mu_1, \mu_2, \mu_3(\rho_1, \rho_2))$ in the ${\mbox{\boldmath${\cal R}$}} = 
\{\rho_1, \rho_2, \Phi\}$ coordinate system, 
where $\mu_3 = \frac{I_1I_2}{I} = \frac{\mu_1\mu_2\rho_1^2\rho_2^2}{\mu_1\rho^2 + \mu_2\rho^2}$.  
Thus I arrive at the action (\ref{ERPMac1}).  

\mbox{ }

\noindent{\bf A.4 Variation of the fully reduced ERPM action}

\mbox{ }

\noindent The momenta are
\beq
{\cal P}_i \equiv \frac{\pa \fL}{\pa \dot{\rho}} = \mu_i\dot{\rho}_i/\dot{\mbox{I}} = \mu_i\rho_i^{\prime} 
\mbox{ } , \mbox{ }  
{\cal P}_3 \equiv {\cal P}_{\Phi} = \frac{\pa \fL}{\pa \dot{\Phi}} = \frac{I_1I_2}{I}\dot{\Phi}/\dot{\mbox{I}} = 
\frac{I_1I_2\Phi^{\prime}}{I}\mbox{ } ,  
\eeq
for $\mbox{}^{\prime} \equiv \frac{\pa}{\pa t} = \frac{1}{\dot{\mbox{\scriptsize I}}}\frac{\pa}{\pa\lambda} = 
\sqrt{\frac{{\sfU + \sfE}}{{\sfT}}}\frac{\pa}{\pa\lambda}$,  where $t$ is the Leibniz--Mach--Barbour 
time \cite{B94I, SemiclI} choice that simplifies the momentum-velocity relations and Euler--Lagrange equations of motion.     
One then discovers as a primary constraint the energy constraint
\beq
{\ttH} \equiv \mbox{}_{{\mbox{\scriptsize{\boldmath${\cal N}$}}}}
\|{\mbox{{\boldmath${\cal P}$}}}\|^2/2 + \fV({\mbox{{\boldmath${\cal R}$}}}) = 
\frac{{\cal P}_1^2}{2\mu_1} + \frac{{\cal P}_2^2}{2\mu_2} + 
\left\{\frac{1}{I_1} + \frac{1}{I_2}\right\}\frac{{\cal P}_{\Phi}^2}{2} + 
\fV(\rho_1, \rho_2, \Phi) = \fE \mbox{ } , \mbox{ } 
\eeq
for ${\mbox{{\boldmath${\cal N}$}}}({\mbox{{\boldmath${\cal R}$}}}) = {\mbox{{\boldmath${\cal M}$}}}^{-1}$ 
the inverse `mass' matrix (which plays an analogous r\^{o}le to that by the DeWitt supermetric 

\noindent
\cite{DeWitt67} in GR), which has components $\mbox{diag}\left(\frac{1}{\mu_1}, \frac{1}{\mu_2}, \frac{1}{\mu_3} \right)$
in the ${\mbox{\boldmath${\cal R}$}} = \{\rho_1, \rho_2, \Phi\}$ coordinate system.

The Euler--Lagrange equations are then 
\beq
\left\{ 
\frac{I_1I_2}{I}\Phi^{\prime}
\right\}^{\prime} = 
\frac{\pa \fU}{\pa \Phi} \mbox{ } ,
\eeq
or, in expanded form, 
\beq
I{\Phi}^{\prime\prime} + 
\left\{ 
\frac{\rho^{\prime}_2}{\rho_2} - \frac{\rho^{\prime}_1}{\rho_1}
\right\}
\{I_1 - I_2\}{\Phi}^{\prime} =
\frac{I^2}{I_1I_2}\frac{\pa \fU}{\pa\Phi}
\mbox{ } ,
\eeq
and
\beq
\mu_i{\rho_i}^{\prime \prime } - \mu_i\rho_i\frac{I_j^2}{I^2}{\Phi^{\prime 2}} = \frac{\pa\fU}{\pa\rho_i}
\eeq
for $(i,j) = (1,2)$ and $(2,1)$.
These propagate the constraint $\ttH$, so the Dirac procedure yields no more constraints.

Note that one can take the Lagrangian form of the energy constraint, 
\beq
\mbox{}_{\mbox{\scriptsize{\boldmath${\cal M}$}}}\|\mbox{\boldmath${\cal R}$} \|^2/2 + 
\fV(\mbox{\boldmath${\cal R}$}) = \frac{\mu_1\rho_1^{\prime 2}}{  2  } + 
\frac{  \mu_2\rho_2^{\prime 2}  }{  2  } + \frac{I_1I_2}{I}\Phi^{\prime 2} + 
\fV(\rho_1, \rho_2, \Phi) = \fE \mbox{ } ,
\eeq
as a first integral in place of one of the three equations of motion.  

\mbox{ }

\noindent{\bf A.5 Simplifications to classical ERPM equations for angle-free potentials}

\mbox{ }

\noindent If $\fV$ is independent of $\Phi$, then $\Phi$ is a cyclic coordinate and the $\Phi$ Euler--Lagrange 
equation simplifies considerably: $\frac{I_1I_2}{I}\Phi^{\prime} = J, \mbox{constant}$, so 
\beq
\Phi^{\prime} = J\left\{\frac{1}{I_1} + \frac{1}{I_2}\right\} 
\mbox{ } .  
\label{angleq}
\eeq
This can be used to remove $\Phi^{\prime}$ from the other equations of motion, 
\beq
\mu_i\rho_i^{\prime\prime} - \frac{J^2}{\mu_i\rho_i^3} = \frac{\pa \fU}{\pa\rho_i} \mbox{ } , 
i = 1 , 2 \mbox{ } ,
\eeq
one of which can also be supplanted by the first integral
\beq
\frac{1}{2}
\sum_{i = 1}^2
\left\{
\mu_i\rho_i^{\prime 2} + \frac{J^2}{\mu_i\rho_i^2}
\right\}
+ \fV(\rho_1, \rho_2) = \fE \mbox{ } .
\eeq
Thus one has ordinary centrifugal terms, but with shared value of angular momentum, corresponding to one 
subsystem having angular momentum $J$ and the other having angular momentum $-J$ (so overall there is 
zero angular momentum).  
I.e. $J$ is a `relative angular momentum quantity'.  

\mbox{ }

\noindent{\bf A.6 Hamilton--Jacobi formulation of ERPM}

\mbox{ }

\noindent The corresponding Hamilton--Jacobi equation is 
\beq
_{\mbox{\scriptsize{\boldmath${\cal N}$}}}||\pa_{\mbox{\scriptsize{\boldmath${\cal R}$}}}\fW||^2/2 
+ \fV(\mbox{\boldmath${\cal R}$}) = \fE
\eeq
for $\fW(\mbox{\boldmath${\cal R}$})$ Hamilton's characteristic function.  
If $\fV = \fV(\mbox{\boldmath$\rho$})$ alone, $\Phi$ is cyclic, so this may be rewritten as 
\beq
_{\bn}||\pa_{\mbox{\scriptsize{\boldmath$\rho$}}}\fW^*||^2/2 + \fV^*(\mbox{\boldmath$\rho$}) = \fE
\label{HJ2} \mbox{ } ,
\eeq
for effective potential $\fV^* = \alpha_{\Phi}^2
\left\{
\frac{1}{I_1(\rho_1)} + \frac{1}{I_2(\rho_2)} 
\right\}
+ \fV(\mbox{\boldmath$\rho$})$, and using $\fW(\mbox{\boldmath${\cal R}$}) = 
\alpha_{\Phi}\Phi + \fW^*(\mbox{\boldmath$\rho$})$.  
If furthermore $\fV(\mbox{\boldmath$\rho$}) = \fV(\rho_1) + \fV(\rho_2)$, (\ref{HJ2}) is separable into 
\beq
\{\pa_{\rho_i}\fW\}^2/\{2\mu_{i}\} + \fV(\rho_{i}) = 
\fE_{i} \mbox{ } ,  
\eeq
for $\fW^*(\mbox{\boldmath$\rho$}) = \fW_1({\rho_1}) + \fW_2({\rho_2})$ and $\fE = \fE_1 + \fE_2$.
Then eqs (\ref{HJ1}--\ref{8}) immediately follow.  


\noindent{\bf{\large Appendix B. Manipulating the SRPM action into fully reduced form}}

\mbox{ }

\noindent{\bf B.1 Variation of the SRPM action (\ref{SRPMAc})}

\mbox{ }

\noindent The momenta are 
\beq
\p^I = \frac{m_I}{I\dot{\mbox{I}}}\delta^{IJ}\vec{{\cal S}}\q_J 
\mbox{ } 
\eeq
for $\dot{\mbox{I}} = \sqrt{\frac{\sfT}{\sfU + \sfE}}$.
As in ERPM, these obey a primary constraint, now of form
\beq
{I}\mbox{ }_{\sbn}||\bp||^2/2 + \fV = \fE
\mbox{ } ,
\eeq
and also free end point variation with respect to $\a$, $\b$ yield the zero momentum and zero angular momentum constraints.  
However now also free end point variation with respect to $\zeta$ yields a zero total dilational momentum constraint, 
\beq
\ttD(\q_I, \p^I) \equiv \sum_{I = 1}^{N} \q_I \cdot \p^I = 0 \mbox{ } .
\eeq
These constraints all propagate, provided that the cofactor of the kinetic energy in the Jacobi action 
is homogeneous of degree 0. 
Moreover, the zero dilational momentum constraint leads straightforwardly to $\dot{I} = 0$, i.e. to the moment of inertia 
being a conserved quantity in this theory.  

\mbox{ }

\noindent{\bf B.2 Elimination of the translations, rotations and dilations}

\mbox{ }

\noindent $\dot{\a}$ and $\dot{\b}$ can again be straightforwardly eliminated from the $\a$ and $\b$ variational 
equations, done in \cite{06II} in terms of $\r_{IJ}$ and here recast in terms of relative Jacobi 
coordinates: 
\beq 
\fI_{\sJacobi}[\R_i,\dot{\R}_i,\dot{\zeta}] = 
\int\d\lambda\fL(\R_i,\dot{\R}_i,\dot{\zeta}) = 2\int\d\lambda\sqrt{\fT^{\$}\{\fU + \fE\}}
\eeq
for
\beq
\fT^{\$}(\R_i,\dot{\R}_i, \dot{\zeta}) = \fT^{\$}_{\sd} + \fT^{\$}_{\sR} \mbox{ } ,
\label{Trel}
\eeq
where 
\beq
\fT^{\$}_{\sd} = 
\mbox{ }_{\mbox{\scriptsize{\boldmath$\mu$}}}||\dot{\bR} + \dot{\zeta}\bR||^2/2I
\eeq
and $\fT^{\$}_{\sR}$ is given by (\ref{fT})/$I$.    
This last numerator is indeed unchanged, for its constituent parts' only velocity dependence is in the 
${\cal L}$ and therein the $\zeta$ corrections cancel by symmetry-antisymmetry.  
One can now eliminate the dilations from the Lagrangian form of the $\zeta$-variation equation, 
$\dot{\zeta} = - E/I$, so $\fT^{\$}_{\sD} \equiv \left.\fT^{\$}_{\sd}
\right|_{\zeta-\mbox{\scriptsize eliminated}} = \frac{IT - E^2}{2I^2}$.   
To recast in relative Jacobi bipolars, use that we have already computed $\fT(\R_i, \dot{\R}_i)$.  
Then, by combining (\ref{fT}) and $I\fT^{\$}(\R_i, \dot{\R}_i) = I\fT^{\$}_{\sD} + \fT_{\sR}$,  
$I\fT^{\$}(\R_i, \dot{\R}_i) = I\fT^{\$}_{\sD} - T/2 + \fT(\R_i, \dot{\R}_i)$, 
so $2I^2\fT^{\$}(\R_i, \dot{\R}_i) = IT - E^2 - IT + 2I\fT(\R_i, \dot{\R}_i) = 
2I\fT(\R_i, \dot{\R}_i) - E^2$, 
\beq
2I^2\fT^{\$}(\rho_i, \dot{\rho}_i, \dot{\Phi}) = IP + I_1I_2\dot{\Phi^2} - E^2 = 
\mu_1\mu_2\{\{\rho_1\dot{\rho}_2 - \rho_2\dot{\rho}_1\}^2 + \rho_1^2\rho_2^2\dot{\Phi}^2\} 
\mbox{ } .
\label{roll}
\eeq

Next introduce the simple ratio shape variable (\ref{SRSV}).  
Then by the quotient rule, (\ref{roll}) becomes 
\beq
2I^2\fT^{\$}(\rho_1, {\cal R}, \dot{{\cal R}}, \dot{\Phi}) = 
I_1^2\{\dot{{\cal R}}^2 + {\cal R}^2\dot{\Phi}^2\} \mbox{ } ,
\eeq
so the fully reduced action takes the form (\ref{redSRPM}).

\mbox{ } 

\noindent{\bf B.3 Variation of the fully reduced SRPM action}

\mbox{ } 

\noindent The momenta are 
\beq
P_{{\cal R}} = {\cal R}^{\prime} \mbox{ } \mbox{ } , \mbox{ } \mbox{ }  
P_{{\Phi}} = {\cal R}^2{\Phi}^{\prime}
\eeq
for $ ^{\prime}$ defined in obvious parallel with App A.4.  
The momenta obey the primary constraint 

\noindent
\beq
{\cal H} = \frac{1}{2}
\left\{
P_{{\cal R}}^2 + \frac{P_{\Phi}^2}{{\cal R}^2}
\right\}  
+ {\cal V}({\cal R}, \Phi) = {\cal E}({\cal R}) \mbox{ } .
\eeq
The Euler--Lagrange equations are 
\beq
{\cal R}^{\prime\prime} - {\cal R}\Phi^{\prime 2} = \frac{\pa \{{\cal U} + {\cal E}\}}{\pa {\cal R}} 
\mbox{ } \mbox{ } , \mbox{ } \mbox{ } 
\eeq
\beq
\{{\cal R}^2\Phi^{\prime}\}^{\prime} = \frac{\pa {\cal U}}{\pa \Phi} \mbox{ } .  
\label{PhiEL}
\eeq
One of these can be supplanted by a first integral 
\beq
\frac{1}{2}
\left\{
{\cal R}^{\prime 2} + {\cal R}^2\Phi^{\prime 2}
\right\} 
+ {\cal V}({\cal R}, \Phi) = {\cal E}({\cal R})
\label{OI}
\eeq
(which is closely related to the primary constraint).  

\mbox{ }

\noindent{\bf{\large Appendix C. Shape--scale coordinate systems for ERPM.}}

\mbox{ }
  
\noindent 

\noindent The shape coordinates used in this paper to study SRPM are also useful in ERPM alongside use of the 
moment of inertia $I$ as a scale variable, both as a neat formulation of dynamics in its own right and 
toward more elaborate explicit toy models of internal time \cite{Kuchar92, Isham93} than those in 
\cite{06II}.    
In this formulation, 
\beq
\fT(I, \Theta, \dot{I}, \dot{\Theta}, \dot{\Phi}) = 
\frac{1}{2}\frac{1}{4I}\{\dot{I}^2 + I^2\{\dot{\Theta}^2 + \mbox{sin}^2\Theta \dot{\Phi}^2\}\}
\mbox{ } .
\eeq
Note that $I$ may be interpreted as a radius.  
Each surface of contant $I (> 0)$ is the shape space, which is topologically $\mathbb{S}^2$.  
Within this picture it is then clear that motions for which $I$ is conserved (as e.g. is always 
the case in SRPM) lie on $\mathbb{S}^2$.       
Also note that the associated metric is conformally a flat space.  
The conformal factor goes singular at the origin (triple collision), corresponding to 
a curvature sigularity in the geometry.  
One could alternatively interpret this as a flat space with a modified potential, 
by refactorizing the corresponding Jacobi action.  
It is then the potential that is singular.

With extension beyond 3 particles in mind, it is thinking not in terms of the above 
$\mathbb{S}^2 \times \mathbb{R}^+$ 
but rather in terms of $\mathbb{CP}^1 \times \mathbb{R}^+$ that is useful.  
In this formulation, then, 
\beq
\fT(I, {\cal Z}, \dot{I}, \dot{{\cal Z}}) = \frac{1}{2}\frac{1}{4I}
\left\{
\dot{I}^2 + I^2\frac{|\dot{{\cal Z}}|^2}{\{ 1 + |{\cal Z}|^2 \}^2}
\right\}
\mbox{ } .
\eeq
%
%


\begin{thebibliography}{99}
\bibitem{AngleDep}           E. Anderson, forthcoming paper with potentials that depend on the relative 
angle treated at the classical level.
\end{thebibliography}
\end{document}